# Size Disorder as a Descriptor for Predicting Reduced Thermal Conductivity in Medium- and High-Entropy Pyrochlores

Andrew J. Wright[a], Qingyang Wang[b], Shu-Ting Ko[c], Ka Man Chung[c], Renkun Chen[b, c], Jian Luo[a, c, *]

[a]Department of NanoEngineering; [b]Department of Mechanical and Aerospace Engineering; [c]Program of Materials Science and Engineering
University of California, San Diego
La Jolla, CA 92093, USA

## Abstract

High-entropy ceramics generally exhibit reduced thermal conductivity, but little is known about what controls this suppression and which descriptor can predict it. Herein, 18 medium- and high-entropy pyrochlores were synthesized to measure their thermal conductivity and Young's modulus. Up to 35% reductions in thermal conductivity were achieved with retained moduli, thereby attaining insulative yet stiff properties for potential thermal barrier coating applications. Notably, the measured thermal conductivity correlates well with a modified size disorder parameter $\delta^*_{Size}$. Thus, this $\delta^*_{Size}$ is suggested as a useful descriptor for designing thermally-insulative medium- and high-entropy ceramics (broadly defined as "compositionally-complex ceramics").

[*] Corresponding should be addressed to J. Luo (jluo@alum.mit.edu).

# Graphical Abstract

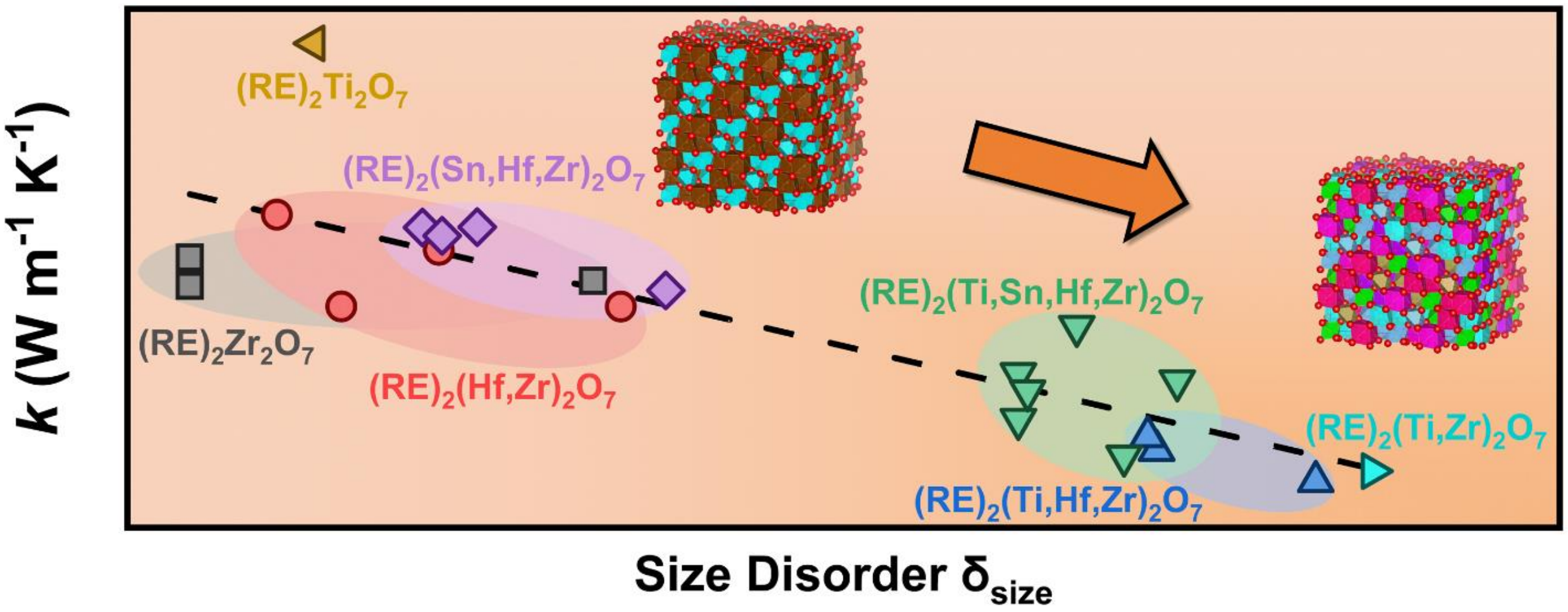

# Highlights

- >20 pyrochlores synthesized, including many new high-entropy compositions
- Mixing of cations at both A and B sites investigated
- All medium- and high-entropy pyrochlores exhibit reduced thermal conductivities
- Thermal conductivity correlates well with a modified size disorder parameter
- A new descriptor proposed for designing thermally-insulative high-entropy ceramics

High-entropy ceramics (HECs), including rocksalt [1], perovskite [2], fluorite [3], and pyrochlore [4,5] oxides, borides [6], carbides [7–10], silicides [11,12] and silicates [13,14] have been reported in the last four years and attracted great research interests. In addition, single-phase high-entropy aluminides (intermetallic compounds), which bridge the well-studied high-entropy alloys (HEAs) [15–17] with new families of HECs, have been made [18]. Some unusual properties of HECs, such as colossal dielectric constants [19], super ionic conductivity [20], and increased hardness [7], have been reported.

One exciting and widely-observed property of HECs is represented by the reduced thermal conductivity [3,10,11,21–24]. While reduced thermal conductivity is generally expected from stronge phonon scattering in HECs, only a few studies have investigated the underlying mechanisms or key controlling factor(s) in depth. Recently, Braun et al. [23] attributed the reduction in thermal conductivity in six-component rocksalt oxides to disorder in the force constants arising from charge variations. Yang et al. [25] also attributed the low thermal conductivity in $Ln_3NbO_7$ (albeit not an HEC) to the force-constant disorder, suggested to be a result of both increased size disorder and charge variation. Furthermore, both reports [23,25] noted high Young's modulus to thermal conductivity ($E/k$) ratios, indicative of a large phonon scattering rate. However, a descriptor to predict the reduced thermal conductivity is still lacking.

To unveil the key parameter controlling the reduced thermal conductivity in HECs, we have fabricated and investigated a series of 22 single-phase pyrochlores, including18 medium- and high-entropy pyrochlores (referred to as "M/HEPs") and four additional low-entropy benchmark compositions. While we recognize there are many definitions/criteria of HEAs [17], here we loosely use "high entropy" to refer to ceramics with five or more cations of high (typically 5%-35%) concentrations (or ideal mixing entropy $\Delta S_X^{mix,ideal}$ >1.5*R* per mol of cations, where *R* is gas constant) on at least one sublattice (if there are multiple ordered cation sublattices), while "medium entropy" refers to cases with three or four cations of high concentrations (or $R < \Delta S_X^{mix,ideal} < 1.5R$) on at least one sublattice. As we will show later, the thermal conductivity is better correlated with size disorder, but not the entropy itself, so that we are not limited to HECs for searching new low-*k* materials. Thus, we propose to extend HECs to "compositionally-complex ceramics (CCCs)" to include both medium-entropy and high-entropy compositions.

With this definition, this study examined 15 MEPs and 3 HEPs, which represent perhaps the largest set of CCCs made and investigated in a single study.

Pyrochlores of $A_2B_2O_7$ structure can be considered as ordered fluorites with two cation sublattices, which represent a potential alternative to yttria-stabilized zirconia (YSZ) as thermal barrier coatings (TBCs) [26]. Cubic pyrochlores have moderately high Young's moduli ($E$ ~ 250 GPa) [27–29] and low thermal conductivities ($k$ ~ 2 W m$^{-1}$ K$^{-1}$) [30–33]. Further efforts to reduce the thermal conductivity include introducing another rare earth element on the A-site [34–41] or B-site [40,41]. Recently, a couple of studies reported the fabrication single-phase high-entropy zirconate-based pyrochlores with low thermal conductivities (albeit possibly some porosity and microstructure effects) [4,5]. Herein, we fabricated >20 dense M/HEPs, where we systematically varied mass and size disorder on both cation sublattices. Subsequently, we identify a useful descriptor for predicting reduced thermal conductivity.

In this work, we adopt the mass disorder ($g$) and size disorder ($\delta$) parameters that are used in thermal transport [42,43] and HEAs [44,45] fields, which are defined as:

$$g = \sum_{i=0}^{n} x_i \left(1 - {m_i}/{\bar{m}}\right)^2 \quad (1)$$

$$\delta = \sqrt{\sum_{i=0}^{n} x_i \left(1 - {r_i}/{\bar{r}}\right)^2} \quad (2)$$

where $x_i$ is the atomic fraction, $m_i$ and $r_i$ are the mass and radius of the i$^{th}$ component, and $\bar{m}$ and $\bar{r}$ are the overall weighted average of mass and ionic radius. Here, we used the Shannon effective ionic radius [46]. Eqs. (1) and (2) assume all cations share the same lattice site; however, the pyrochlore structure has two cation sublattices. Thus, we propose to modify the mass disorder and size disorder parameters as:

$$g^* = \sqrt{g_A^2 + g_B^2} \quad (3)$$

$$\delta_{size}^{*} = \sqrt{\delta_A^2 + \delta_B^2} \tag{4}$$

Here, subscripts "A" and "B" refer to the application of Eqs. (1) and (2) on the A-site and B-site sublattices of $A_2B_2O_7$ pyrochlores (and similar definitions can be extended to other HECs or CCCs with multiple cation sublattices).

All starting constituents were binary oxides purchased from US Research Nanomaterials (~5 μm initial particle sizes). As-received powders were weighed in appropriate amounts, ~2 wt. % stearic acid was added, and the powders were then placed inside a poly(methyl methacrylate) tube with tungsten carbide inserts on both ends along with one Ø5/16" tungsten carbide ball. The powder mixture was dry milled for six hours utilizing a high energy ball mill (SPEX 8000D, SPEX SamplePrep, USA). The milling was interrupted every 30 m to allow for 10 m of cooling to prevent overheating. The milled powders were uniaxially pressed under 100 MPa to form green compacts. The pellets were then placed on a Pt foil inside a calcia-stabilized zirconia crucible within a box furnace and sintered at 1600°C for 24 h.

The densities ($\rho$) were measured following the ASTM C373-18 Standard [47]. The relative densities were calculated from the theoretical densities obtained by X-ray diffraction (XRD). The XRD (Miniflex II, Rigaku, Japan, operating at 30 kV and 15 mA) data were collected at 0.01° 2θ steps. The thermal diffusivity ($\alpha$) was measured by laser flash analysis (LFA 467 *HT HyperFlash*, NETZSCH, Germany). The specific heat capacity ($c_p$) was estimated by Neumann-Kopp law using tabulated values of the constituent oxides [48]. The thermal conductivity ($k$) was then calculated by the product of $\alpha$, $\rho$, and $c_p$ (Suppl. Tables S1). A pulse-echo sonic resonance measurement (TDS 420A, Tektronix, USA) was used to determine the transverse ($u_T$) and longitudinal ($u_L$) waves speeds. From these, Poisson's ratio ($\nu$) and Young's modulus ($E$) were determined by following the ASTM C1198-09 Standard [49].

The relative densities of our sintered specimens were 94-100%. We further corrected the effects of porosity to obtain the intrinsic conductivity and modulus based on [50,51]: $\mathrm{k} = k_{measured}/(1-P)^{3/2}$ and $\mathrm{E} = E_{measured}/(1-2.9P)$, where $P$ is the fraction of pores, and $k$ and $E$ are the corrected intrinsic thermal conductivity and Young's modulus, respectively, of 100% dense specimens that are reported in Table 1 and all figures.

XRD examined the phase formation and crystal structures of all the specimens. Fig. 1 shows XRD pattern, which confirms the formation of the desired pyrochlore phase ($Fd\bar{3}m$, space group #227) without debatable secondary phases, except for Composition P2 and P3 with secondary phases (which were excluded from further measurements).

The measured thermal conductivities and Young's moduli of the referencing $La_2Zr_2O_7$ and $Sm_2Zr_2O_7$ specimens made in this study (Table 1) agree well with prior reports [27,29,37,52]. As shown in Table 1, most M/HEPs exhibited reduced thermal conductivities in comparison with the benchmark $La_2Zr_2O_7$ and $Sm_2Zr_2O_7$ specimens, with up to 35% reductions. Moreover, the reduced thermal conductivity was achieved with retained moduli (all in the range of ~230 – 250 GPa, as shown in Table 1), thereby attaining insulative yet stiff properties (or high $E/k$ ratios). This is an interesting and useful phenomenon for HECs, including high-entropy rocksalts reported earlier [23], as the reduction of thermal conductivity typically occurs along with decreased modulus.

The correlations between three descriptors ($\delta^*_{size}$, $g^*$, $\rho$) and the measured $k$ and $E$ of 22 specimens were examined. In each case, linear regressions were performed, and the degree of correlation was characterized by the Pearson correlation coefficient (PCC) and $r^2$.

Fig. 2 shows that thermal conductivity $k$ and the $E/k$ ratio have the best correlations with the size disorder $\delta^*_{size}$. The size disorder $\delta^*_{size}$ displays a negative correlation (PCC = $-0.86$, $r^2$ = 0.73) with measured $k$. Furthermore, the thermal conductivity $k$ correlates better with the $\delta^*_{size}$, defined in Eq. (4) above, than the individual $\delta_A$ or $\delta_B$ for each sublattice (Suppl. Fig. S1) and the overall $\delta$ assuming random mixing (Suppl Fig. S2). A careful examination further found that the size disorder on the B-site ($\delta_B$) is larger than that on the A-site ($\delta_A$), thereby being more effective in reducing $k$ (Suppl. Fig. S1), being consistent with prior modeling studies [53,54].

A positive correlation was also found between the $E/k$ ratio and $\delta^*_{size}$ (PCC = 0.91, $r^2$ = 0.83). We should emphasize that we do not expect a perfect linear relation ($r^2$ ~ 1) of thermal conductivity $k$ with any single descriptor, since many other factors can influence $k$ (as a complex material property).

Virtually no correlation between $\delta^*_{size}$ and $E$ (PCC = 0.13, $r^2$ = 0.02) was found (Suppl. Fig. S3), implying the high modulus is retained after introducing size disorder in M/HEPs.

The mass disorder $g^*$ was also considered, and its correlations with $k$ and $E/k$ are shown in Suppl. Fig. S4. The mass disorder $g^*$ appears to be a less effective descriptor in comparison with the size disorder $\delta^*_{size}$. Here, the main discrepancies stemmed from the Ti-containing compounds. The correlations of $k$ and $E/k$ with density $\rho$ are weak (Suppl. Fig. S5).

Interestingly, mixing entropies only exhibit weaker correlations with thermal conductivities (Suppl. Fig. S6, albeit there should a correlation between size disorder and mixing entropy, particularly those on the B site). This study showed the size disorder is a stronger descriptor for designing low-$k$ CCCs than the mixing entropy itself.

Additionally, all the compositions containing Ti had $\delta^*_{size}$ > 6% and low $k$ (Fig. 2). As a result, the $E/k$ was significantly higher. Ti is smaller and lighter than other metal elements selected in this study (Suppl. Table S2). In addition to increasing the size disorder, it can serve as a "rattler" on the B-site to reduce $k$ [34]. This effect further shows that ionic radius can be a more significant factor in controlling $k$ than the atomic mass. Consequently, the Ti-containing families exhibit high $E/k$ ratios, achieving 155.6 ± 5.9 GPa·m·K·W$^{-1}$ for Composition P20 $(Sm_{1/3}Eu_{1/3}Gd_{1/3})_2(Ti_{1/4}Sn_{1/4}Hf_{1/4}Zr_{1/4})_2O_7$.

To further explore the Ti effects, the fraction of Ti was systematically varied in $(Sm_{1/3}Eu_{1/3}Gd_{1/3})_2(Ti_xSn_{(1-x)/3}Hf_{(1-x)/3}Zr_{(1-x)/3})_2O_7$. The measured $k$, $E$, and $E/k$ ratio vs. the Ti fraction on the B site ($x$) curves are shown in Fig. 3. It was found that the thermal conductivity $k$ reaches a minimum around $x = 0.25$. The modulus $E$ increases with increasing Ti concentration overall, with a jump when Ti occupies ~50% of the B-site. Consequently, the $E/k$ ratio reaches a maximum of 163.9 ± 6.1 GPa·m·K·W$^{-1}$ for P20-2 $(Sm_{1/3}Eu_{1/3}Gd_{1/3})_2(Ti_{1/2}Sn_{1/6}Hf_{1/6}Zr_{1/6})_2O_7$, where Ti may fit more tightly in the B site to raise $E$.

The rattler effect has been demonstrated in pyrochlores previously [34,40]. Since Yb is small and can behave as a rattler on the A site, a $(Sm_{1/4}Eu_{1/4}Gd_{1/4}Yb_{1/4})_2(Ti_{1/4}Sn_{1/4}Hf_{1/4}Zr_{1/4})_2O_7$ P20 + Yb specimen was synthesized to further verify the rattler effect. The addition of Yb led to a ~13% reduction in $k$ with near negligible loss in $E$ (~4%), thereby resulting in a high $E/k$ ratio of 172.3 ± 5.9 GPa·m·K·W$^{-1}$.

We note that Composition P21 is a material that was recently reported in the literature to have an unusually low RT thermal conductivity of 0.76 W·m$^{-1}$·K$^{-1}$ [4]. However, we measured a much higher RT thermal conductivity of 2.06 $W \cdot m^{-1} \cdot K^{-1}$. Our value appears to be robust since several other HEPs of similar compositions that have similar $k$ values (Table 1). Furthermore, we verified the $k$ values of the two benchmark $La_2Zr_2O_7$ and $Sm_2Zr_2O_7$ specimens widely reported in the literature. In the only other report of HEPs [5], the specimens exhibit similar low $k$ values, but those can be well attributed to the reported low relative densities of ~75%. Thus, the lower $k$ reported in Ref. [4] might also be due to an unknown microstructure effect. Our specimens have high relative densities of 94-100%, and we have corrected the porosity effects to report intrinsic $k$ values in all cases.

In the underlying mechanisms, size disorder can result in both distorted sublattices and oversized cages for the rattler effect, both of which can severely suppress heat conduction [23,25]. Furthermore, unique to the pyrochlore structure, the *8a*-site oxygen ion can relax into the oxygen vacancy site, which likely results in a highly strained anion sublattice [34,40]. Here, the size disorder can be increased in a four-component composition with the largest and smallest elements on each sublattice. To test this effect, Compositions P20-5 and P20-6 were synthesized. On the one hand, P20-6 $(Sm_{3/4}Yb_{1/4})_2(Ti_{1/2}Zr_{1/2})_2O_7$ represents the four-component system with nearly maximized size disorder of 9.50, which exhibits a low $k$ of 1.40 ± 0.04 $W \cdot m^{-1} \cdot K^{-1}$ and a high $E/k$ ratio of 163.1 ± 5.5 $GPa \cdot m \cdot K \cdot W^{-1}$. On the other hand, P20-5 $(Sm_{1/4}Eu_{1/4}Gd_{1/4}Yb_{1/4})_2(Ti_{1/2}Hf_{1/4}Zr_{1/4})_2O_7$, of a slightly lower size disorder of 9.04, shows more attractive properties with the lowest $k$ of 1.36 ± 0.04 $W \cdot m^{-1} \cdot K^{-1}$ and the highest $E/k$ ratio of 175.2 ± 5.7 $GPa \cdot m \cdot K \cdot W^{-1}$ among all 22 specimens examined in this study.

In summary, 22 dense, single-phase cubic pyrochlores, including 18 M/HEPs have been made and measured in this study, representing the largest set of data collected in a single family of HECs and CCCs. Overall, M/HEPs exhibit reduced conductivities (up to ~35%) while retaining similar stiffness (~ 230 – 250 GPa), thereby achieving high $E/k$ ratios, in comparison with simpler pyrochlores such as $La_2Zr_2O_7$, $Sm_2Zr_2O_7$ and $La_2(Hf_{1/2}Zr_{1/2})_2O_7$.

Notably, we find a simple descriptor, size disorder $\delta^*_{size}$ (with a modified definition proposed in this study) to predict reduced thermal conductivities in M/HEPs and potentially other HECs and CCCs. This size disorder parameter $\delta^*_{size}$ is a more effective descriptor to forecast reduce thermal conductivity than the ideal mixing entropy itself. In fact, our three HEPs (P21, P22, and P23) all have relatively higher k values (~2 $W \cdot m^{-1} \cdot K^{-1}$) in comparison with of our best MEPs, *e.g.*, P20 + Yb $(Sm_{1/4}Eu_{1/4}Gd_{1/4}Yb_{1/4})_2(Ti_{1/4}Sn_{1/4}Hf_{1/4}Zr_{1/4})_2O_7$ (~1.39 $W \cdot m^{-1} \cdot K^{-1}$) and P6 $(Sm_{1/2}Gd_{1/2})_2(Ti_{1/3}Hf_{1/3}Zr_{1/3})_2O_7$ (~1.47 $W \cdot m^{-1} \cdot K^{-1}$). This study also supports our proposal to extend HECs to compositionally-complex ceramics or CCCs to include both medium-entropy and high-entropy compositions to search for better properties.

**Acknowledgment:** This material is based upon work supported by the U.S. Department of Energy's Office of Energy Efficiency and Renewable Energy (EERE) under Solar Energy Technologies Office (SETO) Agreement Number EE0008529.

**Table 1.** Summary of the results of all synthesized single-phase pyrochlore oxides.

| Identifier | Composition | $g^*$ (%) | $\delta^*_{Size}$ (%) | $r_A^{3+}/r_B^{4+}$ | Relative Density | $k\ \left(\frac{W}{m \cdot K}\right)$ | $E\ (GPa)$ | $E/k\ \left(\frac{GPa \cdot m \cdot K}{W}\right)$ |
|---|---|---|---|---|---|---|---|---|
| -- | $La_2Zr_2O_7$ | 0.00 | 0.00 | 1.61 | 95.5 ± 0.6% | 2.14 ± 0.06 | 236.2 ± 6.7 | 110.5 ± 4.6 |
| -- | $Sm_2Zr_2O_7$ | 0.00 | 0.00 | 1.50 | 100.0 ± 0.7% | 2.04 ± 0.06 | 220.1 ± 3.0 | 107.9 ± 3.7 |
| P1 | $La_2(Hf_{1/2}Zr_{1/2})_2O_7$ | 10.47 | 0.70 | 1.62 | 98.1 ± 0.6% | 2.29 ± 0.06 | 250.2 ± 4.2 | 109.4 ± 3.6 |
| P4 | $Sm_2(Sn_{1/4}Ti_{1/4}Hf_{1/4}Zr_{1/4})_2O_7$ | 18.86 | 6.65 | 1.58 | 97.7 ± 0.6% | 1.73 ± 0.05 | 250.2 ± 4.7 | 144.3 ± 5.2 |
| P5 | $Gd_2(Sn_{1/4}Ti_{1/4}Hf_{1/4}Zr_{1/4})_2O_7$ | 18.86 | 6.65 | 1.55 | 96.2 ± 0.6% | 1.58 ± 0.05 | 229.4 ± 4.4 | 145.6 ± 5.2 |
| P6 | $(Sm_{1/2}Gd_{1/2})_2(Ti_{1/3}Hf_{1/3}Zr_{1/3})_2O_7$ | 26.33 | 7.76 | 1.57 | 97.9 ± 0.6% | 1.47 ± 0.04 | 242.0 ± 4.2 | 164.8 ± 5.8 |
| P7 | $(Eu_{1/2}Gd_{1/2})_2(Ti_{1/3}Hf_{1/3}Zr_{1/3})_2O_7$ | 26.33 | 7.69 | 1.56 | 97.5 ± 0.6% | 1.52 ± 0.05 | 243.1 ± 4.0 | 160.3 ± 5.6 |
| P8 | $(La_{1/2}Pr_{1/2})_2(Sn_{1/3}Hf_{1/3}Zr_{1/3})_2O_7$ | 7.92 | 2.31 | 1.62 | 94.8 ± 0.6% | 2.24 ± 0.07 | 275.5 ± 7.9 | 122.8 ± 5.1 |
| P9 | $(Eu_{1/2}Gd_{1/2})_2(Sn_{1/3}Hf_{1/3}Zr_{1/3})_2O_7$ | 7.92 | 1.87 | 1.50 | 93.5 ± 0.6% | 2.24 ± 0.07 | 253.3 ± 4.3 | 113.0 ± 3.9 |
| P10 | $(La_{1/3}Pr_{1/3}Nd_{1/3})_2(Hf_{1/2}Zr_{1/2})_2O_7$ | 10.47 | 2.00 | 1.58 | 96.5 ± 0.6% | 2.16 ± 0.06 | 240.0 ± 4.3 | 110.9 ± 3.6 |
| P11 | $(Sm_{1/3}Eu_{1/3}Gd_{1/3})_2(Hf_{1/2}Zr_{1/2})_2O_7$ | 10.47 | 1.22 | 1.49 | 97.9 ± 0.6% | 1.97 ± 0.06 | 235.7 ± 3.8 | 119.8 ± 4.1 |
| P20-1 | $(Sm_{1/3}Eu_{1/3}Gd_{1/3})_2(Sn_{1/3}Hf_{1/3}Zr_{1/3})_2O_7$ | 7.92 | 2.03 | 1.51 | 93.8 ± 0.6% | 2.21 ± 0.07 | 256.7 ± 4.1 | 116.0 ± 4.0 |
| P20 | $(Sm_{1/3}Eu_{1/3}Gd_{1/3})_2(Ti_{1/4}Sn_{1/4}Hf_{1/4}Zr_{1/4})_2O_7$ | 18.86 | 6.73 | 1.56 | 96.4 ± 0.6% | 1.67 ± 0.05 | 260.4 ± 5.0 | 155.6 ± 5.9 |
| P20 + Yb | $(Sm_{1/4}Eu_{1/4}Gd_{1/4}Yb_{1/4})_2(Ti_{1/4}Sn_{1/4}Hf_{1/4}Zr_{1/4})_2O_7$ | 18.87 | 7.50 | 1.54 | 98.1 ± 0.6% | 1.45 ± 0.04 | 249.4 ± 4.0 | 172.3 ± 5.9 |
| P20-2 | $(Sm_{1/3}Eu_{1/3}Gd_{1/3})_2(Ti_{1/2}Sn_{1/6}Hf_{1/6}Zr_{1/6})_2O_7$ | 29.62 | 7.93 | 1.63 | 94.9 ± 0.6% | 1.71 ± 0.05 | 280.1 ± 6.2 | 163.9 ± 6.1 |
| P20-3 | $(Sm_{1/3}Eu_{1/3}Gd_{1/3})_2(Ti_{3/4}Sn_{1/12}Hf_{1/12}Zr_{1/12})_2O_7$ | 33.91 | 7.12 | 1.69 | 94.6 ± 0.6% | 1.90 ± 0.06 | 273.6 ± 5.3 | 144.2 ± 5.2 |
| P20-4 | $(Sm_{1/3}Eu_{1/3}Gd_{1/3})_2Ti_2O_7$ | 0.04 | 1.00 | 1.76 | 94.1 ± 0.6% | 2.88 ± 0.09 | 281.3 ± 5.0 | 97.7 ± 3.4 |
| P20-5 | $(Sm_{1/4}Eu_{1/4}Gd_{1/4}Yb_{1/4})_2(Ti_{1/2}Hf_{1/4}Zr_{1/4})_2O_7$ | 34.07 | 9.04 | 1.58 | 97.8 ± 0.6% | 1.36 ± 0.04 | 238.3 ± 2.5 | 175.2 ± 5.7 |
| P20-6 | $(Sm_{3/4}Yb_{1/4})_2(Ti_{1/2}Zr_{1/2})_2O_7$ | 9.72 | 9.50 | 1.59 | 98.8 ± 0.6% | 1.40 ± 0.04 | 227.7 ± 2.5 | 163.1 ± 5.5 |
| P21 | $(La_{1/5}Ce_{1/5}Nd_{1/5}Sm_{1/5}Eu_{1/5})_2Zr_2O_7$ | 0.13 | 3.24 | 1.54 | 99.1 ± 0.7% | 2.06 ± 0.06 | 230.6 ± 2.9 | 112.1 ± 3.7 |
| P22 | $(La_{1/7}Ce_{1/7}Pr_{1/7}Nd_{1/7}Sm_{1/7}Eu_{1/7}Gd_{1/7})_2(Hf_{1/2}Zr_{1/2})_2O_7$ | 10.47 | 3.46 | 1.55 | 98.6 ± 0.6% | 1.97 ± 0.06 | 242.0 ± 3.5 | 123.0 ± 4.2 |
| P23 | $(La_{1/7}Ce_{1/7}Pr_{1/7}Nd_{1/7}Sm_{1/7}Eu_{1/7}Gd_{1/7})_2(Sn_{1/3}Hf_{1/3}Zr_{1/3})_2O_7$ | 7.92 | 3.82 | 1.56 | 98.7 ± 0.7% | 2.02 ± 0.06 | 187.0 ± 5.0 | 92.4 ± 3.7 |

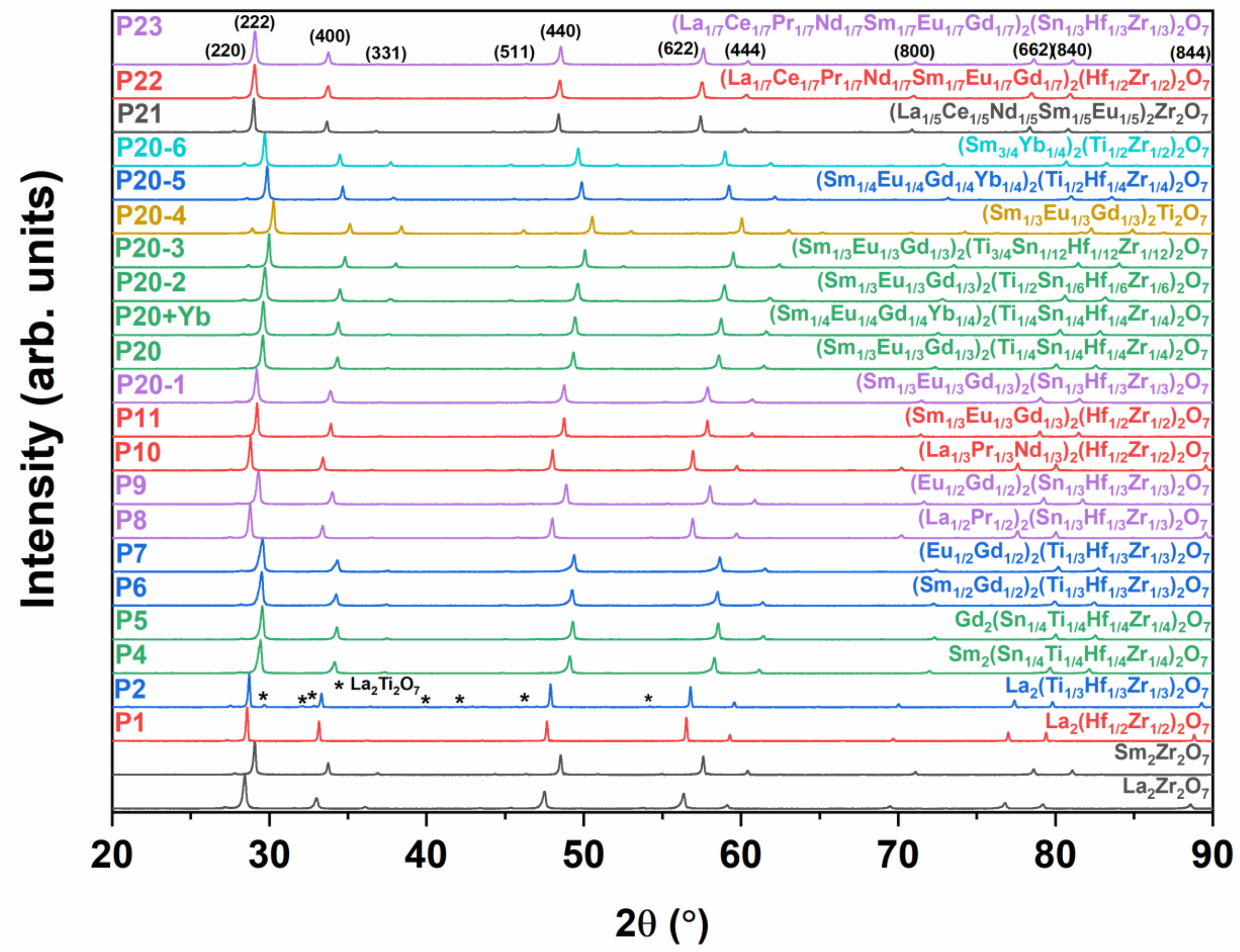

**Fig. 1.** XRD pattern of various pyrochlore oxides synthesized in this study. Composition P2 exhibits secondary phases. The XRD of P3 ($La_2(Sn_{1/4}Ti_{1/4}Hf_{1/4}Zr_{1/4})_2O_7$) is not shown because it was not successfully consolidated into a pellet. The other 22 compositions all exhibit single pyrochlore phases.

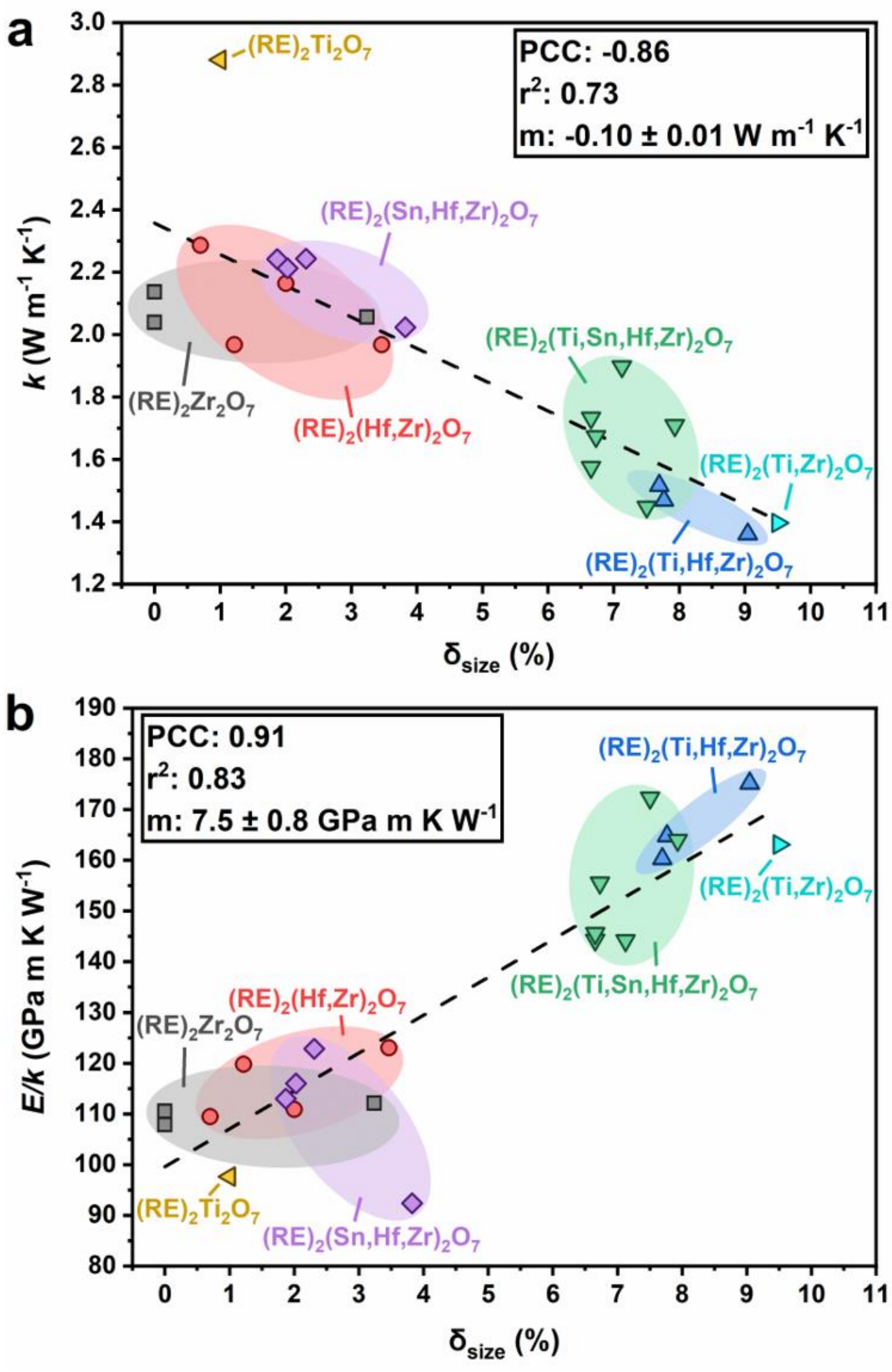

**Fig. 2.** Correlation of (a) thermal conductivity ($k$) and (b) the $E/k$ ratios of all 22 single-phase pyrochlores made in this study with the size disorder parameter, $\delta^*_{size}$. The error bars are given in Table 1, which are not shown here for figure clarity.

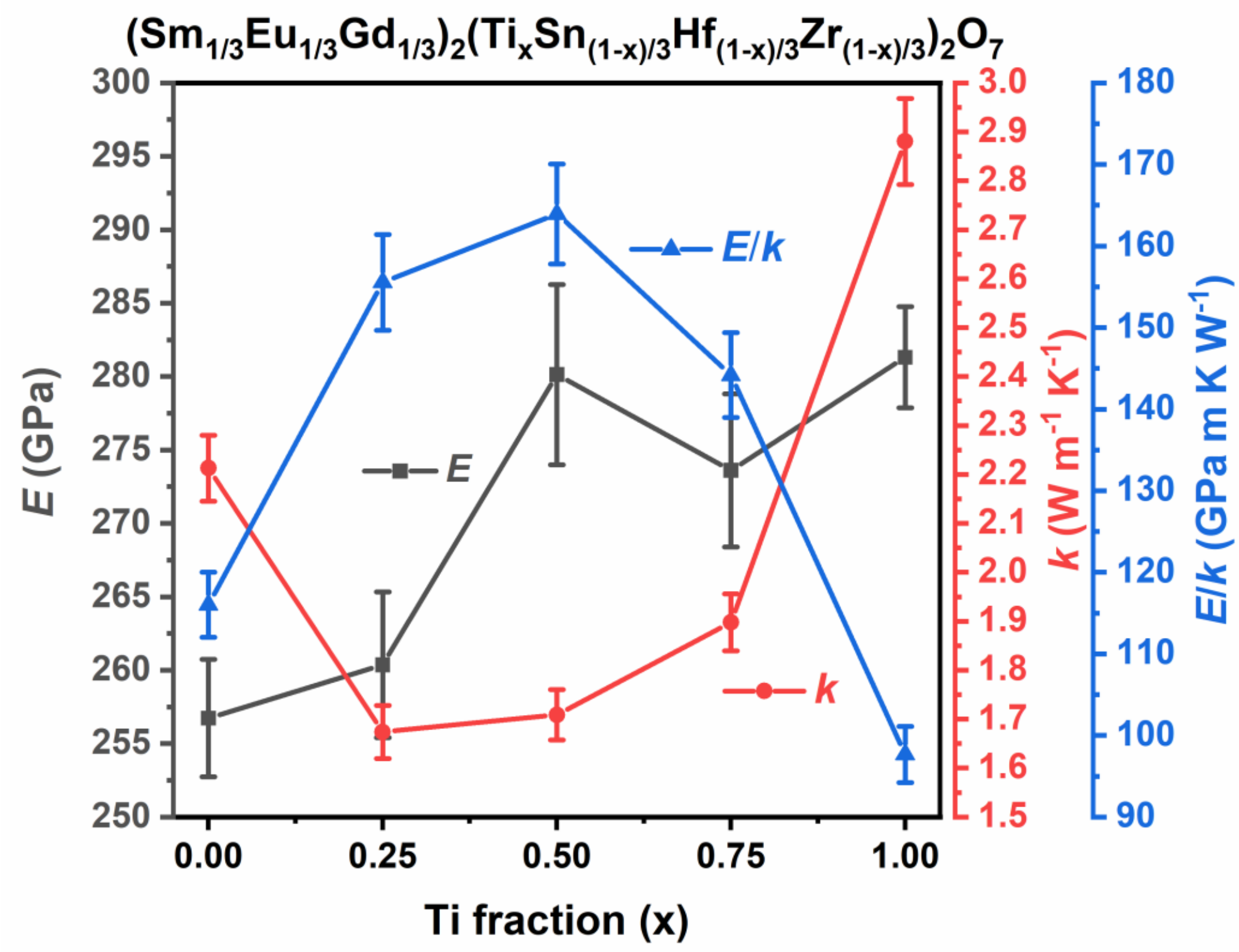

**Fig. 3.** Variations in Young's modulus ($E$), thermal conductivity ($k$), and the $E/k$ ratio as functions of Ti concentration in P20-*n* ($(Sm_{1/3}Eu_{1/3}Gd_{1/3})_2(Ti_xSn_{(1-x)/3}Hf_{(1-x)/3}Zr_{(1-x)/3})_2O_7$).